\documentclass{article} 
\usepackage[accepted]{icml2025}

\usepackage[english]{babel}
\usepackage{algorithmic}
\usepackage{algorithm}
\usepackage{pifont}
\usepackage{tabularx}
\usepackage{amsmath}
\usepackage{bm}
\usepackage{amssymb}
\usepackage{graphicx}
\usepackage{url}
\usepackage{hyperref}
\usepackage[capitalize,noabbrev]{cleveref}
\usepackage{booktabs}
\usepackage{wrapfig}
\usepackage[nolist]{acronym}
\usepackage{tcolorbox}
\usepackage{multirow}
\usepackage{siunitx}
\usepackage{markdown}
\usepackage[normalem]{ulem}
\usepackage{subcaption}
\usepackage{amsmath}
\usepackage{amssymb}
\usepackage{mathtools}
\usepackage{amsthm}

\def\eqref#1{equation~\ref{#1}}

\def\1{\bm{1}}

\def\vp{{\bm{p}}}

\def\vx{{\bm{x}}}

\DeclareMathAlphabet{\mathsfit}{\encodingdefault}{\sfdefault}{m}{sl}
\SetMathAlphabet{\mathsfit}{bold}{\encodingdefault}{\sfdefault}{bx}{n}

\def\gF{{\mathcal{F}}}

\def\gS{{\mathcal{S}}}

\def\gX{{\mathcal{X}}}

\newcommand{\E}{\mathbb{E}}
\newcommand{\Ls}{\mathcal{L}}

\DeclareMathOperator*{\argmin}{arg\,min}

\usepackage{xcolor}
\newcommand{\gray}[1]{\textcolor{gray}{#1}}

\newif\ifcomments
\commentstrue  

\begin{acronym}
  \acro{SD}{stable diffusion}
  \acro{KD}{knowledge distillation}
  \acro{PC}{prompt consistency}
  \acro{ASPKD}{Active Self-Paced Knowledge Distillation}
  \acro{GAN}{Generative Adversarial Network}
\end{acronym}

\theoremstyle{plain}

\theoremstyle{definition}

\theoremstyle{remark}

\icmltitlerunning{Stealix: Model Stealing via Prompt Evolution}

\begin{document}

\twocolumn[
\icmltitle{Stealix: Model Stealing via Prompt Evolution}



\icmlsetsymbol{equal}{*}

\begin{icmlauthorlist}
\icmlauthor{Zhixiong Zhuang}{yyy,comp}
\icmlauthor{Hui-Po Wang}{sch}
\icmlauthor{Maria-Irina Nicolae}{comp}
\icmlauthor{Mario Fritz}{sch}
\end{icmlauthorlist}

\icmlaffiliation{yyy}{Saarland University, Saarbrücken, Germany}
\icmlaffiliation{comp}{Bosch Center for Artificial Intelligence, Renningen, Germany}
\icmlaffiliation{sch}{CISPA Helmholtz Center for Information Security, Saarbrücken, Germany}

\icmlcorrespondingauthor{Zhixiong Zhuang}{zhixiong.zhuang@bosch.com}

\icmlkeywords{Machine Learning, ICML}

\vskip 0.3in
]



\printAffiliationsAndNotice{}

\begin{abstract}
Model stealing poses a significant security risk in machine learning by enabling attackers to replicate a black-box model without access to its training data, thus jeopardizing intellectual property and exposing sensitive information.
Recent methods that use pre-trained diffusion models for data synthesis improve efficiency and performance but rely heavily on manually crafted prompts, limiting automation and scalability, especially for attackers with little expertise.
To assess the risks posed by open-source pre-trained models, we propose a more realistic threat model that eliminates the need for prompt design skills or knowledge of class names.
In this context, we introduce Stealix, the first approach to perform model stealing without predefined prompts. Stealix uses two open-source pre-trained models to infer the victim model’s data distribution, and iteratively refines prompts through a genetic algorithm, progressively improving the precision and diversity of synthetic images.
Our experimental results demonstrate that Stealix significantly outperforms other methods, even those with access to class names or fine-grained prompts, while operating under the same query budget. These findings highlight the scalability of our approach and suggest that the risks posed by pre-trained generative models in model stealing may be greater than previously recognized.\footnote{The project page is at \href{https://zhixiongzh.github.io/stealix/}{https://zhixiongzh.github.io/stealix/}.}

\end{abstract}

\section{Introduction}
\label{sec:introduction}


Model stealing allows attackers to replicate the functionality of machine learning models without direct access to training data or model weights. By querying the victim model with hold-out datasets, the attacker can construct a proxy model that behaves similarly to the original by mimicking its predictions. This attack vector compromises the model owner's intellectual property and may expose sensitive information, posing both security and privacy risks~\citep{dualstudent, pmlr-v235-carlini24a}.

\begin{figure}
    \centering
    \includegraphics[width=0.95\linewidth]{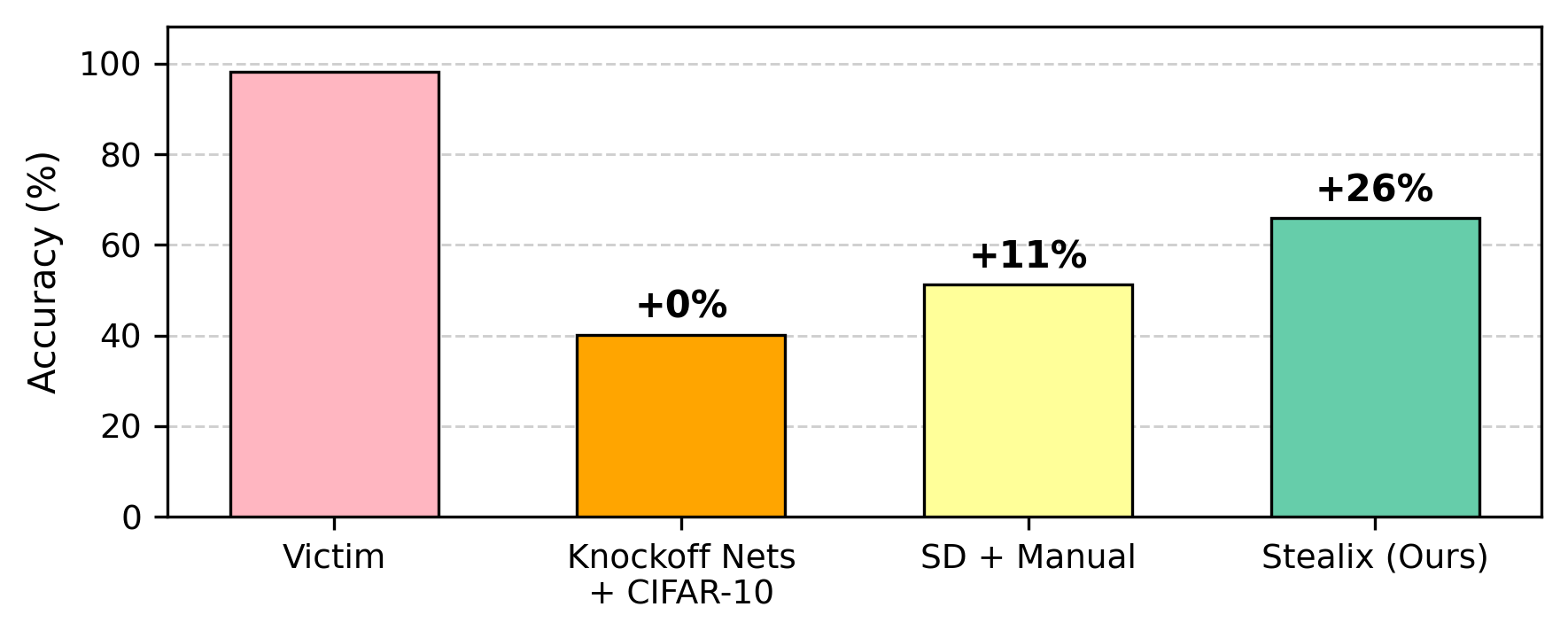}
    \caption{Impact of query datasets on stealing a satellite image classifier: performance drops occur with dissimilar datasets (Knockoff Nets + CIFAR-10) and challenging prompt design (SD + manual). Stealix mitigates these issues by leveraging victim-aware automatic prompt tuning.}
    \label{fig:teaser-comp}
\end{figure}

\begin{figure*}[t!]
    \centering
    
    \includegraphics[width=0.95\textwidth]{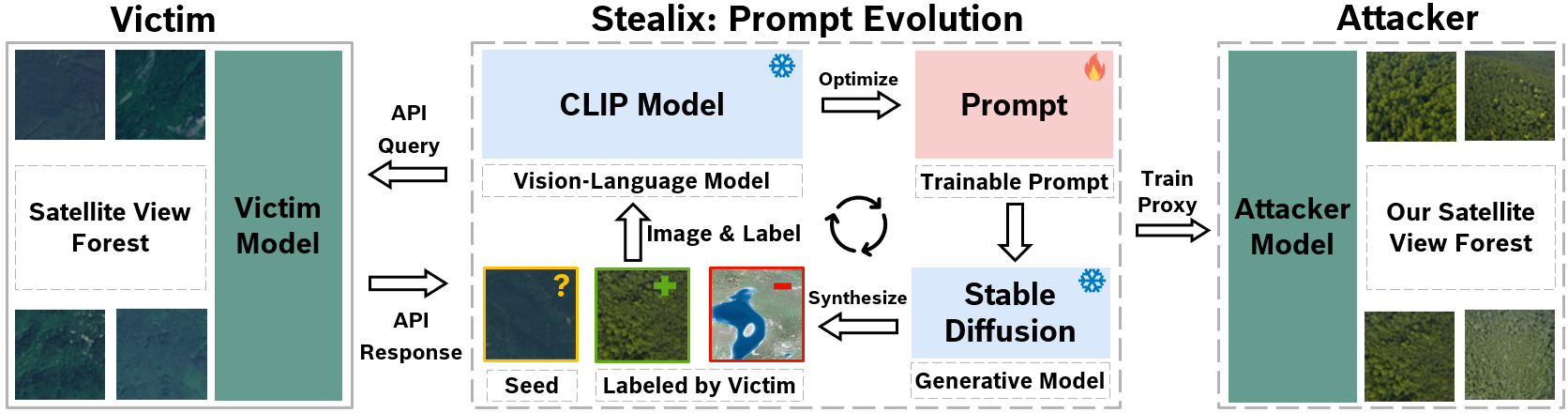}
    \vskip 0.1in
    \caption{Overview of Stealix. Stealix begins with a real image as a seed and synthesizes images to aid model stealing by iteratively refining prompts based on the victim's responses. The synthesized images are then used to train a proxy model.
    }\label{fig:teaser}
\end{figure*}

Current model stealing methods for image classification can be categorized based on the source of the queried images: (1) using publicly available images like Knockoff Nets~\citep{knockoff}, (2) generating images by training a \ac{GAN} from scratch~\citep{dfme, dfme_hardlabel}, or (3) synthesizing images by prompting pre-trained open-source generative models~\citep{shao2023data, hondru2023towards}. The latter uses models like \acf{SD}~\citep{rombach2022high} to achieve superior efficiency by reducing the dependence on online data sources and by eliminating the high computational cost of training new generators.
However, previous approaches often rely on human-crafted prompts or class names to generate images. These methods fall short when the class names lack context or fail to represent the victim’s data features accurately. Attackers may also struggle to describe the target data distribution due to limited knowledge or vague articulation.
Furthermore, reliance on human intervention hinders scalability and automation. These challenges are especially pronounced in specialized fields, where high-value models are the most common. Therefore, research under the current assumptions may oversimplify the problem and underestimate the threat of model stealing facilitated by pre-trained models, as shown in \Cref{fig:teaser-comp}.

To address these limitations and accurately assess the risk, we propose a more realistic threat model in which the attacker lacks prior knowledge or expertise in designing prompts for the victim's data.
This setup reflects practical attack scenarios, such as competitors or malicious actors with limited data but access to black-box model APIs. Under these constraints, existing prompt-based approaches struggle to generate diverse, class-specific queries, limiting their ability to extract the victim model effectively.

In this context, we introduce Stealix, the first model stealing attack that removes the need for human-crafted prompts. 
Our method employs a text-to-image generative model and a vision-language model to iteratively generate multiple refined prompts for each class, as depicted in \Cref{fig:teaser}.
Unlike prior prompt optimization works~\citep{wen2024hard, textual-inversion, da-fusion}, which do not consider the victim model’s predictions in optimizing prompts, our approach incorporates these predictions during the optimization to address inconsistencies in image classification and improve image diversity. We achieve this with contrastive learning and evolutionary algorithms. Specifically, the prompt describing the target class is optimized under a contrastive loss using features extracted by the vision-language model from the prompt itself and from image triplets.
To further improve the precision and diversity of the prompts, we propose a proxy metric as the fitness function to evaluate and evolve the prompts. In practice, our approach requires only a single real image per class.
We show that this is sufficient to achieve new state-of-the-art performance without requiring manual prompt engineering; this assumption is realistic, as potential attackers, typically competitors, often have limited data available, but fail to synthesize more. 

\noindent\textbf{Contributions.}
(i) We present a practical threat model that removes the need for prompt design expertise, reflecting scalability needs in real-world settings.
(ii) We propose Stealix, the first prompt-agnostic approach that iteratively refines prompts using a proxy metric. Statistical analysis demonstrates a high correlation between the proxy metric and the feature distance to the victim data. 
(iii) Stealix surpasses methods using class names or human-crafted prompts, improving attacker model accuracy by up to 22.2\% under a low query budget.
(iv) Our findings reveal critical risks in model stealing with open-source models, underscoring the need for stronger defenses.

\section{Related Works}
\label{sec:relatedworks}
\noindent\textbf{Knowledge distillation.}
\Ac{KD} is a model compression technique that trains smaller student models to replicate the performance of larger teacher models, thereby reducing resource demands~\citep{dodeep?,hintonkd}. 
Traditional \ac{KD} relies on the teacher’s training data to align the student with the same distribution. When this data is unavailable due to practical constraints, surrogate datasets~\citep{datafreekd} or data-free \ac{KD} with generators~\citep{dfad, zeroshotkd} are commonly used, which typically require white-box access for back-propagation. In contrast, model stealing operates in a black-box setting, where the attacker has limited knowledge of the victim model.

\noindent\textbf{Model stealing.} 
Model stealing seeks to replicate a victim model's attributes, such as parameters, hyperparameters~\citep{wang2018stealing, tramer2016stealing}, and functionality~\citep{oliynyk2023know}. Functionality stealing involves training a proxy model to mimic the victim's outputs, raising security concerns in image recognition~\citep{dfme}, natural language processing~\citep{krishna2019thieves}, robotics~\citep{zhuang2024stealthy}, and multimodal radiology report generation~\citep{shen2025medical}. Our work focuses on functionality stealing in images, where traditional methods achieve it by querying victim models using public datasets~\citep{knockoff} or synthetic images~\citep{dfme, dfme_hardlabel, dualstudent}. As illustrated in~\Cref{fig:teaser-comp}, these approaches are either constrained by query dataset similarity or require millions of queries with substantial computational costs.
Recent approaches use pre-trained diffusion models to reduce the query costs~\citep{shao2023data, hondru2023towards}. For instance, \ac{ASPKD}~\citep{hondru2023towards} generates images using diffusion models, queries a subset through the victim model, and pseudo-labels samples via nearest-neighbor matching. However, these methods still depend on class names or manual prompts, limiting their practicality in specialized domains. Our approach introduces automatic prompt refinement to minimize human intervention and thus enhance effectiveness and scalability.


\noindent\textbf{Personalized image synthesis.}
Prompt optimization can capture the essence of specific images, enabling pre-trained text-to-image models to generate personalized outputs.
Textual inversion~\citep{textual-inversion} updates prompt embeddings with text-to-image models, while PEZ~\citep{wen2024hard} optimizes discrete prompts with vision-language model. Notably, DA-Fusion~\citep{da-fusion} leverages textual inversion to synthesize visually similar images for data augmentation. While DA-Fusion is not designed for model stealing, we extend it by replacing the original class label with the victim model’s prediction. Unlike existing approaches, which lack awareness of the victim model’s outputs and generate suboptimal queries, our method explicitly incorporates victim feedback.

\section{Threat Model}
\label{sec:threatmodel}
In this section, we formalize the threat model for model stealing. We start with notations and definitions, describe the victim's capabilities, and outline the attacker's goals and knowledge, emphasizing the constraints that make model stealing challenging.

\paragraph{Notations.}  
Let \( \mathcal{D} = \{(\vx_i, y_i)\} \) be the dataset used to train an image classification model, where \( \vx_i \in \mathbb{R}^{H \times W \times C} \) represents input images with height \( H \), width \( W \), and \( C \) channels, and \( y_i \in \{1, 2, \dots, K\} \) denotes the corresponding class labels, with \( K \) being the total number of classes. Each class is indexed by \( c \in \{1, 2, \dots, K\} \). The pre-trained generative model \( G \) generates an image \( \vx \sim G(\vp, \epsilon) \) from a given prompt \( \vp \) by denoising noise \( \epsilon \) drawn from a standard normal distribution \( \epsilon \sim \mathcal{N}(0, 1) \). For brevity, we denote this process as \( \vx \sim G(\vp) \).

\paragraph{Victim model.}  
The victim trains a classification model \( V \) with parameters \( \theta_v \) on a dataset \( \mathcal{D}_V \), where images are drawn from the victim data distribution \( \vx \sim \mathcal{P}_V \). Once deployed, it operates as a black-box accessible for queries. 
We assume the victim model provides only the top-1 predicted class as answer, thus already reducing the model stealing risks by limiting the attack surface~\citep{dfme_hardlabel}.
For a given input image $\vx$, \( y^* = V(\vx; \theta_v) \in \{1, 2, \dots, K\} \) is the predicted class label.

\paragraph{Goal and knowledge of the attacker.}  
The attacker's objective is to train a surrogate model \( A(\vx; \theta_a) \), parameterized by $\theta_a$ that replicates the behavior of the victim model \( V \). The attacker has black-box access to \( V \), allowing them to query the model with images and receive the predicted class labels. The attacker is constrained by a query budget, representing the total number of queries available per class, denoted as \( B \). The attacker lacks knowledge of (i) the architecture and parameters of \( V \), (ii) the dataset \( \mathcal{D}_V \) used to train \( V \), and (iii) prompt design expertise.
We also limit the use of class names, as they may by chance serve as good prompts; using them would diverge from the assumption that the attacker lacks prompt design expertise.
This constraint significantly limits the attacker from leveraging a generative model for efficient model stealing.


\section{Approach: Stealix}
\label{sec:approach}

\begin{figure*}[t]
    \centering
    \includegraphics[width=0.85\textwidth]{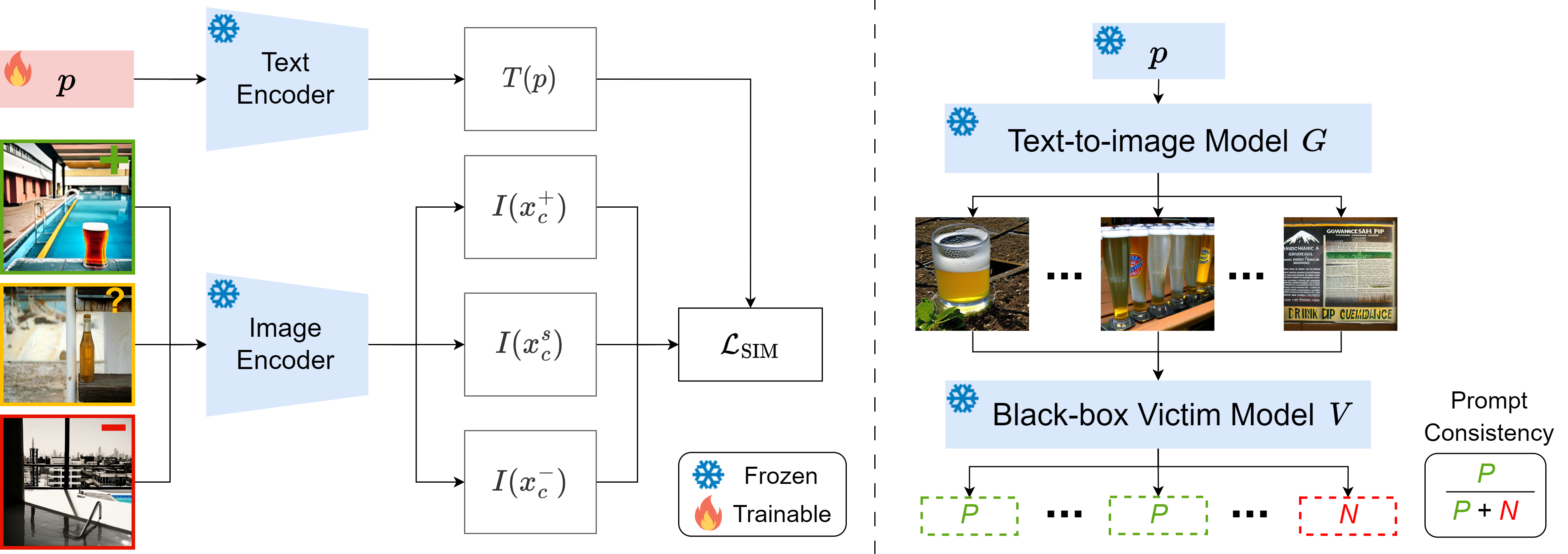}
    \vskip 0.1in
     \caption{Prompt refinement (left) optimizes the prompt $\vp$ using encoders $T$ and $I$ via~\Cref{eq:prompt_refinement} to capture features from seed image $\vx^s$ and positive image $\vx^+$ while filtering out negatives from $\vx^-$. Prompt consistency (right) evaluates $\vp$ with~\Cref{eq:prompt_consistency} by prompting generative model $G$ to synthesize images, which are classified by the victim model $V$ to update positive and negative sets. In the example, the negative feature ``pool'' is removed for class ``bottle''.}\label{fig:method}
            
\end{figure*}


This section details Stealix, formalizing the problem and providing an overview in \Cref{sec:MethodOverview}, followed by explanations of its components in \Cref{sec:PromptRefinement,sec:PromptConsistency,sec:PromptReproduction}.

\subsection{Method Overview}
\label{sec:MethodOverview}
The attacker's goal is to optimize the parameters \( \theta_a \) of a surrogate model \( A \) to replicate the behavior of the victim model \( V \) on the victim data distribution \( \mathcal{P}_V \), achieving comparable performance.
This can be expressed by minimizing the loss between the outputs of the victim and surrogate models over the victim’s data distribution under the cross-entropy loss:%
\begin{equation}
\label{eq:target}
\argmin_{\theta_a} \underset{\vx \sim \mathcal{P}_V}{\E}\left[\Ls_{\text{CE}}\left(V(\vx),A(\vx)\right)\right].
\end{equation}

Without access to the victim data distribution $\mathcal{P}_V$, previous methods~\citep{shao2023data, hondru2023towards} turn to generate high-quality images using a pre-trained text-to-image model \( G \) with a prompt \( \mathbf{p} \). By designing prompts to synthesize images similar to the victim data, the attacker can effectively steal the model by minimizing loss on these generated images:

\begin{equation}
\label{eq:proxy_target}
\argmin_{\theta_a} \underset{\vx \sim G(\vp)}{\E}\left[\Ls_{\text{CE}}\left(V(\vx),A(\vx)\right)\right].
\end{equation}

Recall that for specialized tasks and models, the attacker might be lacking the knowledge to design relevant prompts;
to address this challenge, we propose \textbf{Stealix}.
Through the use of genetic algorithms~\citep{zames1981genetic}, Stealix iteratively generates multiple prompts that capture the diversity of class-specific features recognized by the victim model.

More precisely, each iteration of our attack consists of three steps. \textbf{Prompt refinement} uses a population of image triplets $\mathcal{S}^t$ to optimize corresponding prompts. One randomly initialized prompt is optimized per image triplet to capture the target class features. The resulting prompts are evaluated using \textbf{prompt consistency}, a fitness metric based on how consistently the victim model classifies synthesized images as the target class.
Finally, \textbf{prompt reproduction} evolves the next population of image triplets using a genetic algorithm.
For each iteration $t$, the population $\gS^t = \{ ( \vx_{c}^s, \vx_{c}^+, \vx_{c}^- )_i^t \}_{i=1}^N$, consisting of $N$ image triplets, 
is built using the image sets \( \mathcal{X}_c^s \), \( \mathcal{X}_c^+ \), and \( \mathcal{X}_c^- \), such that $\vx_{c}^s \in \mathcal{X}_c^s$, $\vx_{c}^+ \in \mathcal{X}_c^+$, and $\vx_{c}^- \in \mathcal{X}_c^-$.
These sets are defined for each class $c$: the seed set \( \mathcal{X}_c^s = \{\vx_c^s \mid V(\vx_c^s) = c\} \) contains real images classified as \( c \) by the victim model; the positive set \( \mathcal{X}_c^+ = \{\vx_c^+ \mid V(\vx_c^+) = c\} \) has synthetic images classified as \( c \); and the negative set \( \mathcal{X}_c^- = \{\vx_c^- \mid V(\vx_c^-) \neq c\} \) includes synthetic images classified into other classes than $c$.
$\mathcal{X}_c^+$ and $\mathcal{X}_c^-$ are initially empty, and generated synthetic images are added to these sets over iterations.

The three steps of the method are repeated until the query budget \( B \) per class is exhausted (where \( B = |\mathcal{X}_c^+| + |\mathcal{X}_c^-| \)) (see \Cref{alg:stealix}). Across \( K \) classes, this produces \( K \times B \) synthetic images, which are used along with the seed images to train the attacker model.
We limit the number of seed images the attacker needs to possess from each class to one ($|\mathcal{X}_c^s|=1$). 
The method steps are detailed below.

\subsection{Prompt Refinement}
\label{sec:PromptRefinement}


Efficient model stealing requires synthesizing images that are similar to the victim data, which in turn needs prompts that capture the class-specific features learned by the victim model.
To achieve this, we optimize the prompt to emphasize attributes leading to correct classifications while minimizing misleading features that cause incorrect predictions, with a triplet of images \(\{\vx_c^s, \vx_c^+, \vx_c^-\} \). This triplet, along with a random prompt, is projected into a shared feature space using an image encoder \( I \)  and a text encoder \( T \) from a pre-trained vision-language model (\Cref{fig:method} left). The prompt is then optimized by minimizing the similarity loss between the prompt and image features, with guidance from the victim model’s predictions:


\begin{equation}
\label{eq:prompt_refinement}
\min_\vp \sum_{\vx \in \{\vx_c^s, \vx_c^+, \vx_c^-\}} \mathcal{L}_{\text{SIM}}(I(\vx), T(\vp), V(\vx)),
\end{equation}

where the similarity loss \( \mathcal{L}_{\text{SIM}} \) is defined as:%
\begin{equation}
\label{eq:similarity_loss}
\mathcal{L}_\text{SIM} =
\begin{cases}
    1 - \cos(I(\vx), T(\vp)), & \text{if } V(\vx) = c \\
    \max(0, \cos(I(\vx), T(\vp))), & \text{if } V(\vx) \neq c.
\end{cases}
\end{equation}



If the triplet of images contains only the seed image, the optimization objective degrades to PEZ~\citep{wen2024hard}. We compare ours with PEZ in the ablative study (\Cref{app:ablative}).
This refinement process ensures that the prompt captures salient attributes for accurate classification while eliminating features that may lead to misclassification. See~\Cref{alg:prompt_refinement} in~\Cref{app:alg} for more details.

\subsection{Prompt Consistency}
\label{sec:PromptConsistency}

To evaluate whether the optimized prompt effectively captures the features learned by the victim model, we propose a proxy metric, \acf{PC}.
Since direct access to the victim data distribution is unavailable, this metric serves as an indicator of distribution similarity and is used for prompt reproduction.
We assume that if a prompt captures the latent features of the target class learned by the victim model, the synthetic images will be consistently classified as the target class by the victim model, implying a closer resemblance with the victim data. Based on this assumption, \ac{PC} measures how well a prompt generates images that match the target class \( c \) (\Cref{fig:method} right). Given a prompt \( \vp \), a batch of synthetic images \( \{\vx_i\}_{i=1}^M \sim \mathrm{G}(\vp) \) is generated, where \( M \) is the number of images. \Acl{PC} is computed as:

\begin{equation}
\label{eq:prompt_consistency}
\mathrm{PC} = \frac{1}{M} \sum_{i=1}^M \mathbb{I}(V(\vx_i) = c),
\end{equation}

where \( \mathbb{I}(V(\vx_i) = c) \) is 1 if the victim model classifies \( \vx_i \) as class \( c \), and 0 otherwise. 
In \Cref{subsec:experimental_results}, we show there is a strong correlation between \ac{PC} and the $L_2$ distance between the feature vectors of real and generated images, validating \ac{PC} as an effective proxy metric for monitoring data similarity and for prompt reproduction. The synthetic images are also used to update the image sets $\mathcal{X}_c^+$ and $\mathcal{X}_c^-$, while the \ac{PC} value is added to the fitness set $\gF^t$. Since the prompt is optimized with a triplet of images, the fitness value can also be assigned to the corresponding triplet in $\gS^t$.

\subsection{Prompt Reproduction}
\label{sec:PromptReproduction}

To generate diverse prompts that capture a broad range of class-specific features recognized by the victim model, we evolve the image triplet set $\gS^t$ with $\mathcal{X}_c^s$, $\mathcal{X}_c^+$, and $\mathcal{X}_c^-$ as candidate set. The triplet with the highest fitness value (\ac{PC}) in $\gS^t$ is selected as the elite, carried forward to the next generation $\gS^{t+1}$ to guide the production of improved triplets.
To generate new triplets, $N_p$ triplets are selected from $\gS^t$, where $N_p$ denotes the number of parents. This is done by repeatedly sampling $k$ triplets and selecting the one with the highest fitness to form the parent set $\gS_p$, a process known as tournament selection~\citep{zames1981genetic}, where $k$ is the tournament size. Once the parent set is formed, two parent triplets are selected, and their images are randomly exchanged to create a new triplet.
Each image in the new triplet is replaced with a random sample from $\mathcal{X}_c^s$, $\mathcal{X}_c^+$, or $\mathcal{X}_c^-$ with a probability $p_m$, encouraging exploration of the candidate set. The newly generated triplet is added to $\gS^{t+1}$, and this process is repeated until the population is fully updated. See \Cref{alg:prompt_reproduction} in \Cref{app:alg} for details on the prompt reproduction step.

\begin{algorithm}[t]
\caption{Stealix}
\label{alg:stealix}
\begin{small}
\begin{algorithmic}[1]
   \STATE {\bfseries Input:} seed image set $\{\mathcal{X}_c^s\}_{c=1}^K$, synthetic images amount $M$ for PC calculation, total query budget $B$ per class, population size $N$, victim model $V$, generative model $G$, image encoder $I$ and text encoder $T$
   \STATE {\bfseries Output:} Attacker model $A$
   \STATE Initialize attacker model $A$
   \FOR{each class $c$}
       \STATE $\mathcal{X}_c^+ \gets \emptyset$, $\mathcal{X}_c^- \gets \emptyset$, population index $t \gets 0$, consumed budget $b \gets 0$
       \STATE \textcolor{gray}{// Initial $\gS^0 = \{(x_c^s)_i^0\}_{i=1}^N$ as $\mathcal{X}_c^+, \mathcal{X}_c^-$ are empty.}
       \STATE $\gS^t \gets \{ ( \vx_{c}^s, \vx_{c}^+, \vx_{c}^- )_i^t \}_{i=1}^N $ from $\mathcal{X}_c^s, \mathcal{X}_c^+, \mathcal{X}_c^-$ 
       
       \WHILE{$b < B$}
            \STATE Initialize the fitness score set $\gF^t \gets \emptyset$
            \FOR{each triplet $( \vx_{c}^s, \vx_{c}^+, \vx_{c}^- )_i^t$ in $\gS^t$}
                \IF{$b \geq B$}
                    \STATE \text{break}
                \ENDIF
                \STATE \textcolor{gray}{// Optimize the prompt (\Cref{sec:PromptRefinement})}
                \STATE $\vp_i^t \gets \text{PromptRefinement}(( \vx_{c}^s, \vx_{c}^+, \vx_{c}^- )_i^t, I, T)$
                \STATE \textcolor{gray}{// Synthesize images and get \ac{PC} fitness (\Cref{sec:PromptConsistency})}
                \STATE $\{\vx_{i}\}_{i=1}^M \sim \mathrm{G}(\vp_i^t)$
                \STATE $\gF^t \gets \gF^t \cup \{\frac{1}{M} \sum_{i=1}^M \mathbb{I}(V(\vx_{i}) = c)\}$
                \STATE $b \gets b + M$
                
                \STATE \textcolor{gray}{// Update the positive and negative sets}
                \STATE $\mathcal{X}_c^+ \gets \mathcal{X}_c^+ \cup \{\vx_{i} \mid V(\vx_{i}) = c, \; i \in \{1, \ldots, M\}\}$
                \STATE $\mathcal{X}_c^- \gets \mathcal{X}_c^- \cup \{\vx_{i} \mid V(\vx_{i}) \neq c, \; i \in \{1, \ldots, M\}\}$

            \ENDFOR
        \STATE \textcolor{gray}{// Generate the next population (\Cref{sec:PromptReproduction}).}
            \STATE $\gS^{t+1} \gets \text{PromptReproduction}(\gS^t, \gF^t, \mathcal{X}_c^s, \mathcal{X}_c^+, \mathcal{X}_c^-)$
            
            \STATE $t \gets t + 1$
       \ENDWHILE
   \ENDFOR
   \STATE Train model $A$ with $\{\mathcal{X}_c^+, \mathcal{X}_c^-, \mathcal{X}_c^s\}_{c=1}^K$ and their labels
   \STATE \textbf{return} Attacker model $A$
\end{algorithmic}
\end{small}
\vspace{1pt}
\end{algorithm}

\section{Experiments}
\label{sec:experiments}
In this section, we introduce our experimental results, starting with the experimental setup in \Cref{subsec:experimental_setup}, followed by the results and analyses in \Cref{subsec:experimental_results}. Finally, we exemplify real-world model stealing on a model trained with proprietary data in \Cref{subsec:case_study}.

\subsection{Experimental Setup}

\noindent\textbf{Dataset.}
We train the victim model on four datasets: EuroSAT~\citep{helber2019eurosat}, PASCAL VOC~\citep{Everingham10}, DomainNet~\citep{peng2019moment}, and CIFAR10~\citep{alex2009learning}. Each dataset is chosen for its specific challenges in evaluating model stealing attacks.
EuroSAT requires specialized prompts for satellite-based land use classification, as class names alone fail to generate relevant images. In PASCAL VOC, images are labeled by the largest object, testing the ability to identify the primary target from the victim model's prediction. 
DomainNet evaluates transfer learning across six diverse domains: clipart, infograph, paintings, quickdraw, real images, and sketches. A seed image is randomly chosen from one domain to test cross-domain generalization, using 10 of 345 classes for manageability. In CIFAR10, class names can guide image synthesis, leading to strong baselines when used by other methods, compared to ours, which does not. See~\Cref{app:datasets} for more details.
We also introduce results on two medical datasets in \Cref{app:medical_image_model_stealing}, highlighting the challenges when the diffusion model has limited domain-specific knowledge.

\noindent\textbf{Victim model.}
All models use ResNet-34 following~\citet{dfme}, with PASCAL using an ImageNet-pretrained weights. Victim models are trained with SGD, Nesterov with momentum 0.9, a 0.01 learning rate, $5 \times 10^{-4}$ weight decay, and cosine annealing for 50 epochs.

\noindent\textbf{Stealix.}
We use OpenCLIP-ViT/H as the vision-language model~\citep{cherti2022reproducible} for prompt refinement, with a learning rate of 0.1 and 500 optimization steps using the AdamW optimizer. We employ Stable Diffusion-v2~\citep{rombach2022high} as the generative model, with a guidance scale of 9 and 25 inference steps. \ac{PC} evaluation uses $M=10$ images. Stable Diffusion-v2 is used across all methods. 
In prompt reproduction, we set the population size to \(N = 10\), with \(N_p = 5\) parents selected via tournament selection with a tournament size of \(k = 5\), and retain one elite per generation. The mutation probability is set to \(p_m = 0.6\).
Following prior work~\citep{dfme}, we use ResNet-18 as the attacker model. 
To focus on the impact of query data quality and ensure a fair comparison across methods, we train the attacker model using the same hyperparameters as the victim model without tuning: 50 epochs with SGD.
More attacker and victim architectures are shown in~\Cref{app:attacker_architecture} and~\Cref{app:victim_architecture}.
The experiments are run on a NVIDIA V100 GPU and an AMD EPYC 7543 32-Core CPU. The computation time is provided in~\Cref{app:time_consumption}.


\noindent\textbf{Baselines.}
We focus on a new, practical threat model that lacks both prompt expertise and detailed class information. Nevertheless, we compare our method with existing approaches designed for other threat models.
Specifically, we consider the following baselines. (i) \textbf{DA-Fusion}~\citep{da-fusion} is adapted to train a soft prompt from the seed image using textual inversion, then synthesize query images with strength 1 and the same guidance scale as our method; (ii) \textbf{Real Guidance}~\citep{he2022synthetic} uses the prompt ``a photo of a \{class name\}'' to synthesize images given the seed image with strength 1 and same guidance scale; (iii) \textbf{ASPKD}~\citep{hondru2023towards} follows a three-stage process, first generating 5000 images per class using Real Guidance, then querying the victim model with a limited budget \( B \), and finally pseudo-labeling the remaining images with a nearest neighbors approach with the attacker model;
(iv) \textbf{Knockoff Nets}~\citep{knockoff} evaluates performance with randomly collected images by querying the CIFAR-10 victim model with EuroSAT images and other victim models with CIFAR-10; 
(v) \textbf{DFME}~\citep{dfme} is a data-free model stealing method based on \acp{GAN} that trains a generator from scratch to adversarially generate samples to query the victim model. We report the result of DFME using a query budget of 2 million per class.
(vi) \textbf{\ac{KD}}~\citep{hintonkd} serves as a reference upper bound, where the attacker queries the victim model using its training data to evaluate the best possible performance with data access.
While data augmentation without querying the victim model is not model stealing, we include a comparison of attacker model accuracy between model stealing and data augmentation setups in~\Cref{app:model_stealing_outperforms_data_augmentation}.

\noindent\textbf{Evaluation metrics.} 
We rely on two metrics: (i) the accuracy of the attacker model on the test set of the victim data, which is standard for stealing classifiers~\citep{knockoff}, and (ii) the \acf{PC} of the synthesized images. For Stealix, we report the best \ac{PC} achieved across varying query budgets. For Real Guidance and DA-Fusion, where the prompt remains fixed, \ac{PC} is measured by querying 500 images per class. For \ac{ASPKD} that uses images synthesized by Real Guidance, \ac{PC} is identical to Real Guidance. \ac{PC} is not applicable for \ac{KD}, Knockoff, and DFME, which do not involve text-to-image synthesis. All experiments are repeated three times, with mean values in the table and confidence intervals in the figure.

\label{subsec:experimental_setup}

\subsection{Experimental Results}

\begin{table*}[t]
\caption{Attacker model accuracy with a query budget of 500 per class; DFME uses 2M.}
\label{tab:performance-table}
\vskip 0.1in
\addtolength{\tabcolsep}{-0.2em}
\begin{center}
\begin{small}
{\fontsize{9}{10}\selectfont  
\begin{tabular}{ccccccc}
\toprule
\multicolumn{1}{c}{\bf Method} & \multicolumn{1}{c}{\bf \#Seed images}& \multicolumn{1}{c}{\bf Class name}  & \multicolumn{1}{c}{\bf EuroSAT} & \multicolumn{1}{c}{\bf PASCAL} & \multicolumn{1}{c}{\bf CIFAR10} & \multicolumn{1}{c}{\bf DomainNet} \\ \midrule

Victim        & - & - & 98.2\% \gray{(1.00x)} & 82.7\% \gray{(1.00x)}  & 93.8\% \gray{(1.00x)}  & 83.9\% \gray{(1.00x)}  \\

(KD)            & - & - & 95.6\% \gray{(0.97x)} & 34.6\% \gray{(0.42x)} & 76.7\% \gray{(0.82x)} & 74.6\% \gray{(0.89x)} \\ \midrule

Knockoff      & 0 & \ding{55} & 40.1\% \gray{(0.41x)} & 22.3\% \gray{(0.27x)} & 24.4\% \gray{(0.26x)} & 39.3\% \gray{(0.47x)} \\
DFME      & 0 & \ding{55} & 11.1\% \gray{(0.11x)} & 6.6\% \gray{(0.08x)} & 23.7\% \gray{(0.25x)} & 18.5\% \gray{(0.22x)} \\
ASPKD         & 0 & \checkmark & 39.2\% \gray{(0.40x)} & 25.7\% \gray{(0.31x)} & 27.1\% \gray{(0.29x)} & 27.3\% \gray{(0.32x)} \\

Real Guidance & 1 & \checkmark & 51.2\% \gray{(0.52x)} & 24.0\% \gray{(0.29x)} & 27.4\% \gray{(0.29x)} & 31.9\% \gray{(0.38x)} \\

DA-Fusion     & 1 & \ding{55} & 59.0\% \gray{(0.60x)} & 16.4\% \gray{(0.20x)} & 26.7\% \gray{(0.28x)} & 28.4\% \gray{(0.34x)} \\

Stealix (ours)    & 1 & \ding{55} & \textbf{65.9\%} \gray{(0.67x)} & \textbf{40.0\%} \gray{(0.48x)} & \textbf{49.6\%} \gray{(0.53x)} & \textbf{48.4\%} \gray{(0.58x)} \\
\bottomrule
\end{tabular}
}
\end{small}
\end{center}
\end{table*}

\begin{figure*}[t!]
    \begin{center}
    \includegraphics[width=0.85\textwidth]{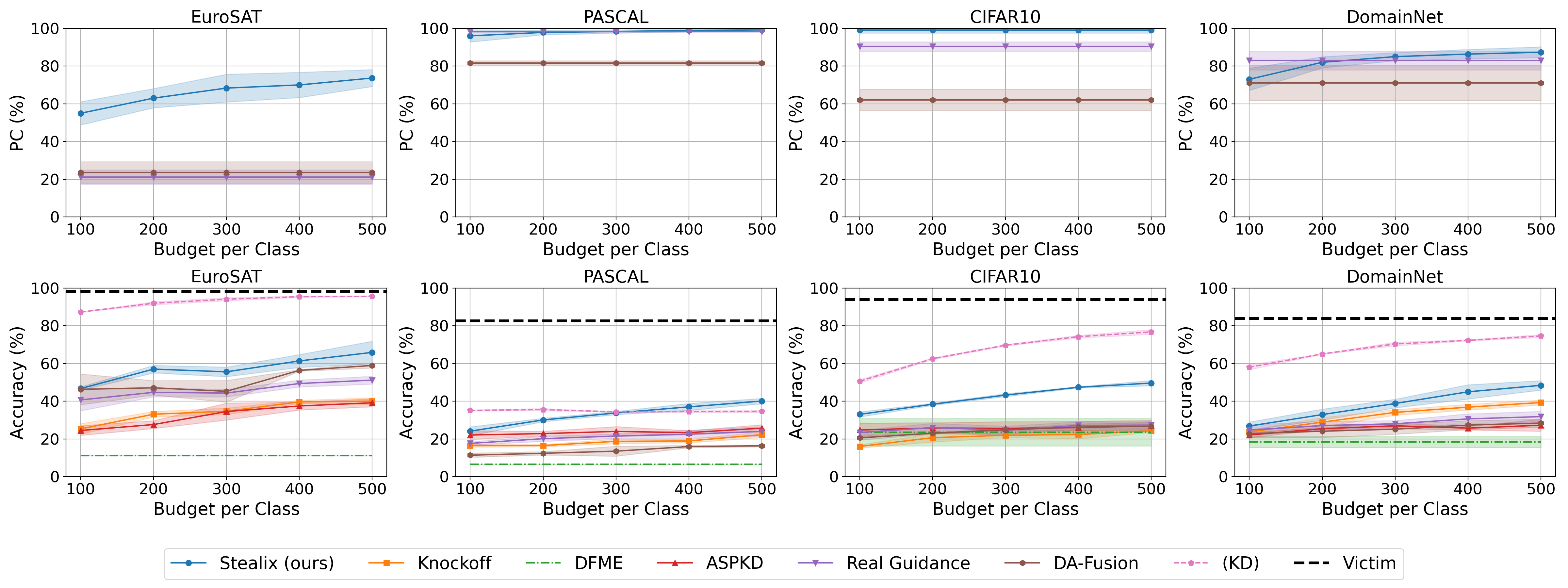}
    \vskip 0.1in
    \caption{PC and attacker model accuracy comparison across datasets with varying query budgets per class. DFME uses 2M per class. Besides the baselines, we provide KD and victim accuracy for reference.}
    \label{fig:result_with_different_budget}
    \end{center}
\end{figure*}

\noindent\textbf{Comparison with baselines.}
\Cref{tab:performance-table} compares the accuracy of the attacker model across methods for a query budget of 500 per class (2M per class for DFME). Stealix consistently outperforms all other methods. E.g., in CIFAR-10, Stealix achieves 49.6\% accuracy, a 22.2\% improvement over the second-best method, Real Guidance. In contrast, DFME has near-random accuracy on EuroSAT and PASCAL due to its reliance on training a generator from scratch with small perturbations, which are quantized when interacting with real-world victim APIs (discussed in~\Cref{app:limitatios_of_dfme}). In PASCAL, Stealix even surpasses \ac{KD}, where the attacker has access to the victim data. This is because \ac{KD} is constrained by the limited victim data size (around 73 images per class), whereas Stealix generates additional images and issues more queries.
In \Cref{fig:result_with_different_budget} we illustrate both the stolen model accuracy and \ac{PC} across varying query budgets. Stealix consistently achieves higher \ac{PC} as the query budget increases, particularly in EuroSAT, where class names alone are insufficient for generating relevant images. Although Real Guidance initially attains higher \ac{PC} in PASCAL and DomainNet, Stealix ultimately surpasses it with larger query budgets. In CIFAR-10, Stealix reaches nearly 100\% \ac{PC}.
In \Cref{app:attacker_architecture} and \Cref{app:victim_architecture}, our method consistently outperforms others with different attacker and victim architectures.


\noindent\textbf{Limitations of human-crafted prompts.}
Even when attackers can craft prompts for the seed image based on the prior knowledge of class names, these prompts, though logically accurate from a human perspective, may still fail to capture the nuanced features learned by the victim model. To evaluate this, we utilize InstructBLIP~\citep{instructblip}, a pre-trained vision-language model, as a proxy for a human attacker. InstructBLIP is instructed with, ``It is a photo of a \{class name\}. Give me a prompt to synthesize similar images,'' alongside the seed image from the challenging EuroSAT dataset. 
We synthesize 500 images per class based on these prompts and train the attacker model. The comparison of generated prompts between InstructBLIP and Stealix for all classes is provided in~\Cref{app:instructblip}, along with examples of generated images.
Stealix outperforms InstructBLIP, achieving an accuracy of 65.9\% compared to 55.2\%. Despite InstructBLIP incorporating relevant terms like ``aerial view'' and ``satellite view,'' its average \ac{PC} score is 41.0\%, compared to Stealix’s 73.7\%.







\begin{figure*}[ht!]
    \begin{center}
    \includegraphics[width=0.85\textwidth]{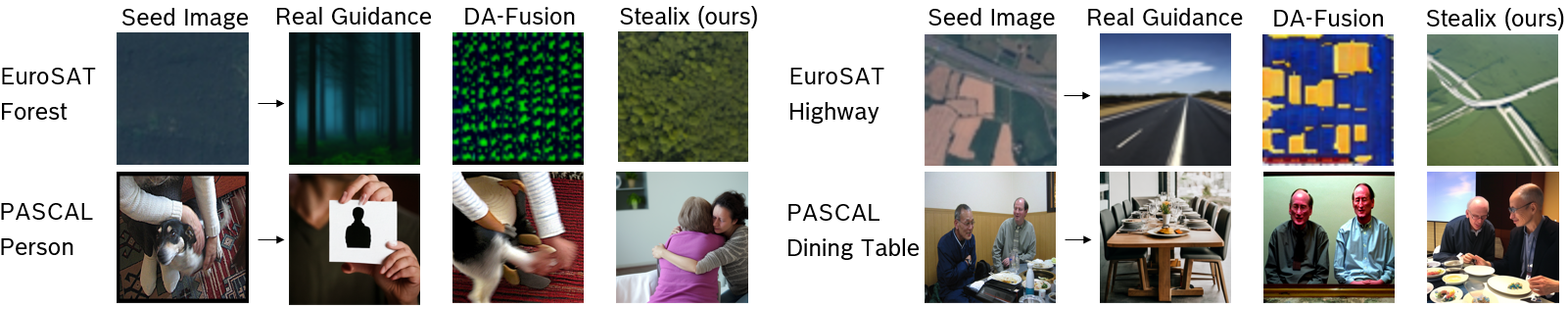}
    \vskip 0.1in
    \caption{Qualitative comparison of images generated by Real Guidance, DA-Fusion, and Stealix.}
    \label{fig:qualitative}
    \end{center}
\end{figure*}

\noindent\textbf{Qualitative comparison.}
\Cref{fig:qualitative} presents qualitative comparisons on EuroSAT and PASCAL datasets. In EuroSAT, class names alone miss attributes like the satellite view, leading Real Guidance to generate generic images that differ from the victim data.
Additionally, DA-Fusion struggles to interpret blurred seed images, generating random color blocks.
For PASCAL, when multiple objects are present in the seed image, Stealix successfully identifies the target object. 
For instance, the seed image for the ``PASCAL Person'' class includes a prominent dog, leading to the first-generation prompt, ``chilean vaw breton cecilia hands console redux woodpecker northwestern \textbf{beagle} sytracker \textbf{collie} relaxing celticsped'', which generates dog images and results in \acl{PC} of 0. Stealix then uses the misclassified image as a negative example and refines the prompt to, ``syrian helene pasquspock hands thumbcuddling sheffield stuck smritihouseholds vulnerable kerswednesday humormindy intestin'', removing dog-related features and achieving \ac{PC} = 1. 
Similarly, Stealix correctly identifies the dining table as the target in a crowded scene, while DA-Fusion incorrectly focuses on the human. These examples show how Stealix evolves prompts by filtering out misleading features using victim feedback.

\noindent\textbf{Correlation between \ac{PC} and feature distance.} 
Since the attacker lacks access to the distribution of the victim data, \ac{PC} is proposed as a proxy for monitoring and optimizing prompts, based on the hypothesis that more consistent predictions from the victim model indicate a closer match to its data. To evaluate this assumption, we collect 150 \ac{PC} values per class corresponding to different prompts during prompt evolution.
For each \ac{PC}, we compute the $L_2$ distance between the mean feature vector of the synthetic images and that of the victim data.
Feature vectors are extracted from the victim model before its final fully connected layer. The Spearman’s rank correlation test shows a strong, statistically significant negative correlation between \ac{PC} and $L_2$ (\Cref{tab:hypothesis}), supporting the use of \ac{PC} as guiding metric. We also evaluate whether higher \ac{PC} leads to higher attacker model performance with different \ac{PC} values in \Cref{app:pc_to_acc}.

\begin{table}[t]
\caption{Spearman's rank correlation between \ac{PC} and $L_2$ feature distance.}
\vskip 0.1in
\label{tab:hypothesis}
\begin{center}
\begin{small}
\begin{tabular}{c   S[table-format=-1.2] S[table-format=1.2e-3]}
\toprule
\textbf{Data} & \textbf{Correlation}~\boldmath$\rho$ & \textbf{p-value} \\
\midrule
EuroSAT & -0.63 & 7.04e-05 \\
PASCAL & -0.64 & 2.79e-04 \\
CIFAR10 & -0.73 & 1.20e-07 \\
DomainNet & -0.88 & 1.83e-26 \\
\bottomrule
\end{tabular}
\end{small}
\end{center}
\end{table}

\noindent\textbf{Diversity comparison.}
\Cref{fig:result_with_different_budget} shows that although \ac{PC} values of Real Guidance are similar to ours for CIFAR10, PASCAL and DomainNet, our attacker model performs consistently better. This advantage stems from the greater diversity in our synthetic data, achieved through prompt evolution, where distinct images are used to construct different triplets. To quantify this, we use the diversity score proposed by~\citet{improved_recall}, Recall, which measures the likelihood that a random image from the victim data distribution falls within the support of the synthetic image set. The higher the score, the more diverse the images. As shown in \Cref{tab:diversity}, our method generates more diverse synthetic data with higher Recall score.

\begin{table}[t]
\caption{Diversity (recall) across methods that using text-to-image generative models; higher scores indicate better diversity relative to the victim data distribution.}
\vskip 0.1in
\label{tab:diversity}
\begin{center}
\begin{small}
\setlength{\tabcolsep}{3.3pt} 
\begin{tabular}{ccccc}
\toprule
\multicolumn{1}{c}{\bf Method} & \multicolumn{1}{c}{\bf EuroSAT} & \multicolumn{1}{c}{\bf PASCAL} & \multicolumn{1}{c}{\bf CIFAR10} & \multicolumn{1}{c}{\bf DomainNet} \\ \midrule
Real Guidance & 0.29 & 0.07 & 0.40 & 0.41 \\
DA-Fusion     & 0.43 & 0.06 & 0.48 & 0.24 \\
Stealix (ours)    & \textbf{0.49} & \textbf{0.30} & \textbf{0.76} & \textbf{0.66} \\
\bottomrule
\end{tabular}
\end{small}
\end{center}
\end{table}

\label{subsec:experimental_results}

\subsection{Stealing Model Based on Proprietary Data}
We now apply Stealix to a large-scale Vision Transformer (ViT)~\citep{dosovitskiy2020vit} trained on proprietary and non-public data, significantly differing from our previous victims. This model is a `Not Safe For Work' (NSFW) binary classification model, publicly available from HuggingFace~\citep{nsfw_model}, and ranked among the top-4 most downloaded models for image classification. 
We use a publicly available NSFW dataset from HuggingFace~\citep{nsfw_data}\footnote{Warning: This dataset contains sexual content. Viewer discretion is advised.} to run this attack.
The dataset contains 200 images (100 ``safe'', 100 ``not safe'').
The victim reaches 92.0\% accuracy on this data.
The attack is initiated with one random image per class, the same hyperparameters from \Cref{subsec:experimental_setup} and a ResNet-18 attacker.
With a query budget of 500 queries per class, Stealix achieves an accuracy of 73.0\%, effectively replicating the victim model.
In contrast, the Real Guidance method fails to synthesize ``not safe for work'' images, resulting in an attacker model accuracy of 50.0\%, equivalent to random guessing. DA-Fusion demonstrates moderate performance, with 62.3\% accuracy.
This result demonstrates that our approach can leverage general priors in diffusion models to enhance model stealing, even in the absence of diffusion models trained on specific datasets.
\label{subsec:case_study}

\section{Discussion}
\label{sec:discussion}
\noindent\textbf{Defense.}
Our threat model assumes that the victim employs a defense that returns only hard label outputs, which is cheap and effective in limiting information leakage compared to soft labels~\citep{dfme_hardlabel}.
\Cref{app:soft_label} shows that the attacker models' accuracy improves with soft-label access using images generated by Stealix, underscoring the need for this defense. 
Similarly, previous works~\citep{lee2019defending, mazeika2022steer} propose defenses that perturb the posterior prediction to reduce the utility of stolen models, while keeping the predicted class (argmax) unchanged to preserve original performance for benign users. These approaches implicitly push attackers to rely on hard labels, which are less informative but immune to such perturbations. However, since our prompt evolution uses only hard label feedback, this constraint impacts only the training of the attacker model, not the optimization of prompts, suggesting that stronger defenses may be required.

\noindent\textbf{Limitations and future work.}
Unlike GAN-based methods, Stealix does not require backpropagation through the victim model to train the generator, which enhances generalization across victim architectures (\Cref{app:victim_architecture}). Although the attacker architecture can still influence the performance (\Cref{app:attacker_architecture}), our method consistently outperforms the baselines. Since image synthesis and surrogate training are decoupled, attackers can reuse synthetic images for, e.g., hyperparameter tuning. This key advantage could be explored in future work to improve model accuracy. Finally, as open-source generative models advance, integrating more powerful models into our framework offers significant potential for further enhancements.


\section{Conclusion}

We showed that attackers can leverage open-source generative models to steal proprietary ones, even without prompt expertise or class information. Using Stealix for prompt evolution and aligning generated data with victim data significantly boosts model extraction efficiency. This is the first study to reveal the risks of publicly available pre-trained generative models in model theft for realistic attack scenarios. We urge the development of defenses against this emerging threat.

\section*{Impact Statement}
This work aims to raise awareness of the risks associated with model stealing, particularly through the use of open-source pre-trained generative models. While our work demonstrates how such models can be exploited in adversarial settings, it is intended to inform the development of more robust defenses against model theft. We emphasize that our approach is not designed to promote malicious behavior but to highlight vulnerabilities that need addressing within the AI community. We encourage practitioners, model developers, and stakeholders to implement stronger defenses, such as hard-label-only responses or adversarial detection mechanisms, to mitigate potential risks. All experiments were conducted with publicly available models and data, and with the intent of advancing the security of AI systems.

\section*{Acknowledgements}
We acknowledge the support and funding by Bosch AIShield. This work was also partially funded by ELSA – European Lighthouse on Secure and Safe AI funded by the European Union under grant agreement No. 101070617, as well as the German Federal Ministry of Education and Research (BMBF) under the grant AIgenCY (16KIS2012).

\bibliography{my_bib}
\bibliographystyle{icml2025}

\newpage
\appendix
\onecolumn

\section{Algorithms}
\label{app:alg}

We detail the algorithms for prompt refinement and prompt reproduction in~\Cref{sec:PromptRefinement} and~\Cref{sec:PromptReproduction}.

\noindent\textbf{Prompt refinement.} We implement the hard prompt optimization method proposed by PEZ~\citep{wen2024hard} to optimize the prompt to capture target class features learnt by the victim model (\Cref{alg:prompt_refinement}). The soft prompt, $\bm{\hat{p}}$, consists of $L$ embedding vectors and is initialized from the vocabulary embedding set $\mathbf{E}$. The soft prompt is iteratively mapped to its nearest neighbor embeddings using a projection function, $\text{Proj}\mathbf{E}(\bm{\hat{p}})$, and converted into a hard prompt, $\vp$, via a function $\text{Soft2Hard}(\bm{\hat{p}})$. During each iteration, the soft prompt is updated through gradient descent, guided by the similarity loss $\Ls_{\text{SIM}}$, which aims to retain features in the positive image while reducing features in the negative image. This process is repeated for $s$ optimization steps, after which the final hard prompt is obtained. We follow the hyperparameters from~\citet{wen2024hard}, setting $L=16$ and $\gamma=0.1$, while reducing $s$ from 5000 to 500 to save optimization time, e.g., on EuroSAT, from approximately 3 minutes to 18 seconds. We further evaluate the impact of prompt lengths (4, 16, 32) on EuroSAT with a query budget of 500 per class across three random seeds. The results show that Stealix achieves 62.5\%, 65.9\%, and 64.3\% accuracy for prompt lengths 4, 16, and 32, respectively. This demonstrates that Stealix consistently outperforms others (second-best method: 59.0\% from DA-Fusion in~\Cref{tab:performance-table}), with prompt length 16 striking the best balance between efficiency and accuracy.

\begin{algorithm}[h]
\caption{Prompt Refinment}
\label{alg:prompt_refinement}

\begin{algorithmic}[1]
   \STATE {\bfseries Input:} image triplet $(\vx_{c}^s, \vx_{c}^+, \vx_{c}^- )$, text encoder $T$ and image encoder $I$, optimization steps $s$, learning rate $\gamma$, soft prompt length $L$
   \STATE {\bfseries Output:} hard prompt $\vp$
   \STATE Initialize soft prompt $\bm{\hat{p}}$ from vocabulary $\mathbf{E}$

    \FOR{step $= 1$ to $s$} 
        \STATE \textcolor{gray}{// Project soft prompt to nearest neighbor embeddings and convert to hard prompt.}
        \STATE $ \bm{\hat{p}}' \gets \text{Proj}_\mathbf{E}(\bm{\hat{p}})$
        \STATE $ \vp \gets \text{Soft2Hard}(\bm{\hat{p}}')$
        
        \STATE \textcolor{gray}{// Compute gradient of the similarity loss and update soft prompt using gradient descent.}
        \STATE $g \gets \nabla_{\bm{\hat{p}}'} \sum_{\mathbf{x} \in (\vx_{c}^s, \vx_{c}^+, \vx_{c}^- )} \mathcal{L}_{\text{SIM}}(I(\vx), T(\vp), V(\vx))$
        \STATE $\bm{\hat{p}} \gets \bm{\hat{p}} - \gamma g$
    \ENDFOR
    
    \STATE \textcolor{gray}{// Final projection to ensure the soft prompt is fully converted to hard tokens.}
    \STATE $ \bm{\hat{p}}' \gets \text{Proj}_\mathbf{E}(\bm{\hat{p}})$
    \STATE $ \vp \gets \text{Soft2Hard}(\bm{\hat{p}}')$
    
    \STATE \textbf{return} hard prompt $\vp$
\end{algorithmic}
\end{algorithm}


\noindent\textbf{Prompt reproduction.} In~\Cref{alg:prompt_reproduction}, we employ a genetic algorithm to iteratively refine prompts through tournament selection, crossover and mutation. In tournament selection, we use prompt consistency as the fitness function.

\begin{algorithm}[h]
\caption{Prompt Reproduction}
\label{alg:prompt_reproduction}
\begin{small}
\begin{algorithmic}[1]
    \STATE {\bfseries Input:} Current population $\gS^t$, fitness set $\gF^t$, seed image set $\mathcal{X}_c^s$, positive image set $\mathcal{X}_c^+$, negative image set $\mathcal{X}_c^-$, tournament size $k$, number of parents $N_p$, number of populations $N$.
    \STATE {\bfseries Output:} Evolved population $\gS^{t+1}$
    
    \STATE Select the elite triplet $(x_c^s, x_c^+, x_c^-)_{\text{elite}}$ with the highest fitness from $\gS^t$ given $\gF^t$
    \STATE Initialize next population $\gS^{t+1} \gets \{(x_c^s, x_c^+, x_c^-)_{\text{elite}}\}$ \textcolor{gray}{// Keep the elite triplet in the next population}
    \STATE Initialize the parents set $\gS_p \gets \emptyset$
    
    \STATE \textcolor{gray}{// Perform tournament selection to select $N_p$ parents.}
    \FOR{$i = 1$ to $N_p$}
        \STATE Randomly select $k$ triplets from $\gS^t$
        \STATE Choose the triplet $(x_c^s, x_c^+, x_c^-)$ with maximum fitness from the $k$ triplets given $\gF^t$
        \STATE $\gS_p \gets \gS_p \cup \{(x_c^s, x_c^+, x_c^-)\}$
    \ENDFOR
    
    \STATE \textcolor{gray}{// Generate the next generation.}
    \FOR{$i = 1$ to $N-1$}
        \STATE \textcolor{gray}{// Apply crossover using selected parents.}
        \STATE Select two parents from $\gS_p$ cyclically, denoted as $(x_{c,1}^s, x_{c,1}^+, x_{c,1}^-)$ and $(x_{c,2}^s, x_{c,2}^+, x_{c,2}^-)$
        \STATE Split each parent at a random point and form a new triplet, e.g., $(x_{c,1}^s, x_{c,1}^+, x_{c,2}^-)$, as new triplet $(x_c^s, x_c^+, x_c^-)$
        \STATE \textcolor{gray}{// Apply mutation.}
        \STATE Replace each image in $(x_c^s, x_c^+, x_c^-)$ with a random one from $\gX_c^s$, $\gX_c^+$, or $\gX_c^-$ with probability $p_m$
        \STATE $\gS^{t+1} \gets \gS^{t+1} \cup \{ (x_c^s, x_c^+, x_c^-)\}$
    \ENDFOR

    \STATE $\gS^{t+1}$
\end{algorithmic}
\end{small}
\end{algorithm}

\section{Datasets}
\label{app:datasets}
We provide an overview of the datasets introduced in our experiment setup (\Cref{subsec:experimental_setup}), detailing the sizes of the training and validation sets and their respective image resolutions (see~\Cref{tab:overview_dataset}). For CIFAR-10, we utilize the standard training and test splits provided by PyTorch, which consist of 50,000 training images and 10,000 test images at a resolution of $32 \times 32$ pixels. In the case of PASCAL, we follow the preprocess from DA-Fusion~\citep{da-fusion} to assign classification labels based on the largest object present in each image, resulting in 1,464 training images and 1,449 validation images with an image size of $256 \times 256$ pixels. The EuroSAT dataset is split into training and validation sets using an 80/20 ratio while maintaining class distribution through stratified sampling, yielding 21,600 training images and 5,400 validation images at a resolution of $64 \times 64$ pixels. For DomainNet, we select the first 10 classes in alphabetical order across six diverse domains: clipart, infograph, paintings, quickdraw, real images, and sketches. We apply the same 80/20 stratified split as used for EuroSAT, resulting in 11,449 training images and 2,863 validation images, each resized to $64 \times 64$ pixels.

\begin{table}[h]
\vskip 0.1in
\caption{Overview of datasets.}
\label{tab:overview_dataset}
\vskip 0.1in
\begin{center}
\begin{tabular}{@{}cccc@{}}
\toprule
\textbf{Dataset} & \textbf{Train/Val} & \textbf{Image Size}  \\ \midrule
EuroSAT & 21.6K/5.4K & $64 \times 64$   \\
PASCAL & 1464/1449 & $256 \times 256$  \\
CIFAR10 & 50K/10K & $32 \times 32$  \\ 
DomainNet & 11449/2863 & $64 \times 64$  \\ 
\bottomrule
\end{tabular}%
\end{center}
\end{table}

\section{Comparison of Computation Time}
\label{app:time_consumption}
We present a comparison of the time required for various methods using the EuroSAT dataset as an example. All experiments were conducted on a single machine with an NVIDIA V100 32GB GPU and an AMD EPYC 7543 32-Core Processor. \Cref{tab:time_and_accuracy} summarizes the total time for the process under a 500-query budget per class (with DFME using 2M queries per class). Stealix demonstrates state-of-the-art accuracy while maintaining reasonable computational efficiency.

\begin{table}[h]
\caption{Comparison of computational time and accuracy across methods on the EuroSAT dataset. The victim model accuracy 98.2\%.}
\label{tab:time_and_accuracy}
\vskip 0.1in
\begin{center}
\begin{tabular}{@{}lcccccc@{}}
\toprule
                 & \textbf{Knockoff} & \textbf{DFME} & \textbf{ASPKD} & \textbf{Real Guidance} & \textbf{DA-Fusion} & \textbf{Stealix (ours)} \\ \midrule
\textbf{Time (hours)}   & 0.5             & 4.5           & 28.6          & 3.3                   & 5.4               & 6.3                   \\
\textbf{Accuracy}  & 40.1\%            & 11.1\%          & 39.2\%          & 51.2\%                  & 59.0\%              & 65.9\%                  \\ \bottomrule
\end{tabular}
\end{center}
\end{table}

\newpage
\section{Ablative Analysis}
\label{app:ablative}
We evaluate the contribution of prompt reproduction to Stealix by conducting an ablation study, where prompts are optimized using only CLIP from the seed image, without reproduction. This degrades our method to a version equivalent to PEZ~\citep{wen2024hard}, which relies solely on a single image and does not consider the victim model's predictions.
More specifically, PEZ optimizes prompts using only the seed image $x_c^s$, while our prompt refinement reformulates prompt optimization as a contrastive loss over a triplet $(x_c^s, x_c^+, x_c^-)$, guided by the victim model’s predictions. This enables Stealix to capture class-relevant features more effectively. Stealix further introduce prompt consistency as a proxy for evaluation, and prompt reproduction using genetic algorithms, forming a complete and victim-aware model stealing framework.
As shown in \Cref{tab:ablation_study}, the setup labeled ``Stealix w/o reproduction (PEZ)'' shows a significant accuracy drop, highlighting the critical role of our victim-aware prompt optimization to evolve the prompts with prompt consistency. Note that PEZ is not originally a model stealing method, but a prompt tuning technique. We include it as part of an ablation study instead of a baseline comparison to isolate the impact of our proposed components.



\begin{table}[h!]
\caption{Ablation study: comparison of attacker model accuracy without prompt reproduction.}
\label{tab:ablation_study}
\vskip 0.1in
\begin{center}
\begin{small}
\begin{tabular}{@{}ccccccc@{}}
\toprule
\multicolumn{1}{c}{\bf Method}  & \multicolumn{1}{c}{\bf EuroSAT} & \multicolumn{1}{c}{\bf PASCAL} & \multicolumn{1}{c}{\bf CIFAR10} & \multicolumn{1}{c}{\bf DomainNet} \\ \midrule

Victim      & 98.2\% \gray{(1.00x)} & 82.7\% \gray{(1.00x)}  & 93.8\% \gray{(1.00x)}  & 83.9\% \gray{(1.00x)}  \\ \midrule
Stealix w/o reproduction (PEZ)   & 60.1\% \gray{(0.61x)} & 26.7\% \gray{(0.32x)} & 33.8\% \gray{(0.36x)} & 39.2\% \gray{(0.47x)} \\
Stealix (ours)  & \textbf{65.9\%} \gray{(0.67x)} & \textbf{40.0\%} \gray{(0.48x)} & \textbf{49.6\%} \gray{(0.53x)} & \textbf{48.4\%} \gray{(0.58x)} \\
\bottomrule
\end{tabular}%
\end{small}
\end{center}
\end{table}

\section{Linking Prompt Consistency to Model Accuracy}
\label{app:pc_to_acc}
\begin{figure}[h!]
    \begin{center}
    \includegraphics[width=1\textwidth]{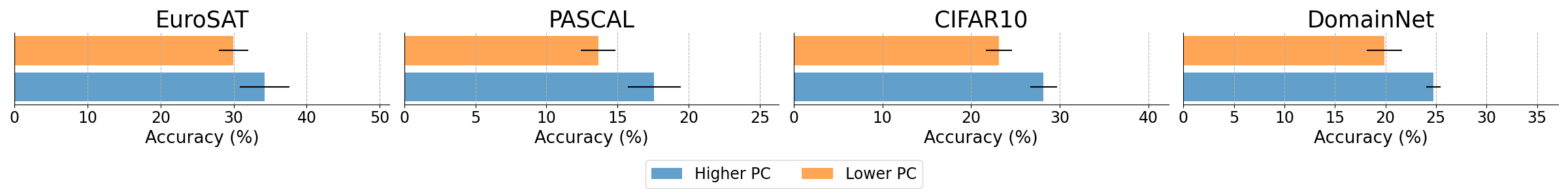}
    \vskip 0.1in
    \caption{Comparison of attacker model accuracy using synthetic images generated from prompts with higher and lower \acl{PC} across four datasets.}
    \label{fig:pc_acc}
    \end{center}
\end{figure}

To evaluate whether higher \ac{PC} leads to more effective model stealing, we compare attacker model performance using synthetic images generated from two prompts with different \ac{PC} values. Specifically, we select prompts at the 25th percentile (lower \ac{PC}) and the 100th percentile (higher \ac{PC}) during the prompt evolution process. We generate 500 synthetic images with each of the two prompts, query the victim model, and use only the positive images to train the attacker model. We exclude the 0th percentile prompt because it yields no positive samples. Since the higher \ac{PC} prompt generates more positive images than the lower \ac{PC} prompt, we reduce the number of positive images from the higher \ac{PC} prompt to match that of the lower \ac{PC} prompt. The results, presented in~\Cref{fig:pc_acc}, demonstrate that higher \ac{PC} values consistently lead to improved attacker model accuracy across all datasets, confirming that higher \ac{PC} enhances the effectiveness of model stealing attacks.

\newpage
\section{Simulating Attacker with InstructBLIP}
\label{app:instructblip}
\begin{figure}[ht!]

    \begin{center}
    \includegraphics[width=1.0\textwidth]{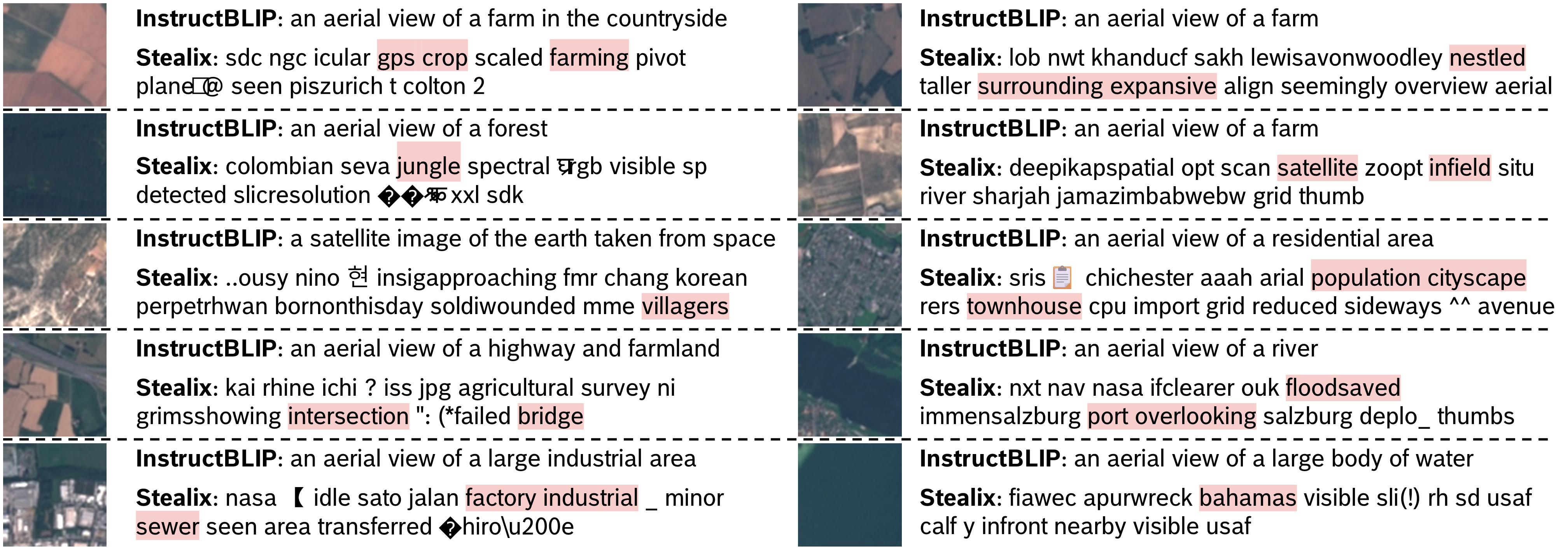}
    \vskip 0.1in
    \caption{Seed images and corresponding prompts generated by InstructBLIP and Stealix for the EuroSAT dataset. Each pair shows the original seed image and the prompt used for image synthesis. Class names from top to bottom, left to right: AnnualCrop, Forest, HerbaceousVegetation, Highway, Industrial, Pasture, PermanentCrop, Residential, River, SeaLake. Feature words related to each class are highlighted in red for Stealix.}
    \label{fig:high_pc_prompt}
    \vskip 0.1in
     \end{center}
\end{figure}

\begin{figure}[ht!]

    \begin{center}
    \includegraphics[width=0.85\textwidth]{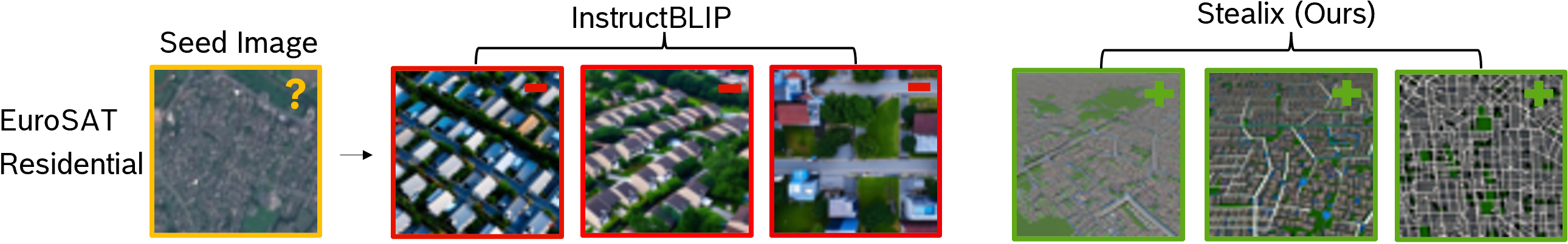}
    \vskip 0.1in
    \caption{Synthetic images for the Residential class with the prompt from InstructBLIP and ours.}
    \label{fig:qualitative_instructblip}
    \vskip 0.1in
     \end{center}
\end{figure}

The prompts generated by InstructBLIP~\citep{instructblip} for the EuroSAT dataset are conditioned on seed images and the instruction: ``It is a photo of a \{class name\}. Give me a prompt to synthesize similar images.'' 
In~\Cref{fig:high_pc_prompt}, we show the prompts produced by InstructBLIP and by Stealix with high \ac{PC}. As discussed in~\Cref{subsec:experimental_results} and shown in~\Cref{tab:instructblip}, prompts generated by InstructBLIP result in lower \ac{PC} values and reduced attacker model performance due to misalignment with the latent features learned by the victim model, despite being human-readable.
For example, in the ``Residential'' class of EuroSAT (\Cref{fig:qualitative_instructblip}), InstructBLIP’s prompt ``an aerial view of a residential area'' results in a
\ac{PC} of only 8.8\%, while Stealix reaches 71.0\%.

As for Stealix, the optimized prompts are not always interpretable to humans, echoing our motivation that human-crafted prompts may be suboptimal for model performance. Moreover, Stealix supplements class-specific details that may be overlooked by humans. For example, as shown in~\Cref{fig:high_pc_prompt} (highlighted in red), \textbf{gps crop} emphasizes geospatial context for AnnualCrop, \textbf{jungle} suggests dense vegetation for Forest, and \textbf{floodsaved, port, and bahamas} convey water-related cues for River and SeaLake. These examples illustrate how Stealix uncovers latent features that the victim learns and highlight the limitations of human-centric prompt design and the importance of automated prompt evolution in model stealing.

        

\begin{table}[ht!]
\caption{Comparison with InstructBLIP on EuroSAT at a query budget of 500 per class.}
\label{tab:instructblip}
\vskip 0.1in
\begin{center}
\begin{tabular}{ccccc}
\toprule
\multicolumn{1}{c}{\bf Method} & \multicolumn{1}{c}{\bf \#Seed images}& \multicolumn{1}{c}{\bf Class name}  & \multicolumn{1}{c}{\bf PC} & \multicolumn{1}{c}{\bf Accuracy} \\ \midrule

InstructBLIP    & 1 & \checkmark & 41.0\% & 55.2\% \\
Stealix (ours)    & 1 & \ding{55} & \textbf{73.7\%}  & \textbf{65.9\%}\\
\bottomrule
\end{tabular}
\end{center}
\end{table}

\newpage
\section{Different Attacker Model Architectures}
\label{app:attacker_architecture}
We analyze the performance of different attacker model architectures, including ResNet18, ResNet34, VGG16, and MobileNet, as shown in~\Cref{tab:attacker_architecture}. Our method, Stealix, consistently outperforms all other baselines, regardless of the attacker model architecture. However, the choice of architecture does impact performance: smaller models like MobileNet result in lower accuracy due to their limited capacity, as seen in the KD baseline where MobileNet achieves only 89.2\% accuracy compared to 95.6\% with ResNet. This suggests that architectural limitations, rather than the attack method, drive the performance drop. Moreover, because Stealix decouples image synthesis from attacker model training, attackers can optimize hyperparameters and architectures without re-querying the victim model, offering flexibility and efficiency.

\begin{table}[ht]
\caption{Performance comparison of different attacker architectures against a ResNet34 victim model (98.2\% accuracy) trained on EuroSAT, using a query budget of 500 queries per class.}
\vskip 0.1in
\addtolength{\tabcolsep}{-0.2em}
\label{tab:attacker_architecture}
\begin{center}
{\fontsize{8}{9}\selectfont
\begin{tabular}{ccc|cccc}
\toprule
\multirow{2}{*}{\bf Method} & \multirow{2}{*}{\bf \#Seed images} & \multirow{2}{*}{\bf Class name} & \multicolumn{4}{c}{\bf Attacker architecture} \\ \cmidrule{4-7}
 &  &  & \multicolumn{1}{c}{\bf ResNet18} & \multicolumn{1}{c}{\bf ResNet34} & \multicolumn{1}{c}{\bf VGG16} & \multicolumn{1}{c}{\bf MobileNet} \\ \midrule

KD            & - & - & 95.6\% \gray{(0.97x)} & 95.6\% \gray{(0.97x)} & 95.7\% \gray{(0.97x)} & 89.2\% \gray{(0.91x)} \\ \midrule

Knockoff      & 0 & \ding{55} & 40.1\% \gray{(0.41x)} & 40.3\% \gray{(0.41x)} & 40.1\% \gray{(0.41x)} & 29.3\% \gray{(0.30x)} \\
DFME          & 0 & \ding{55} & 11.1\% \gray{(0.11x)} & 11.1\% \gray{(0.11x)} & 11.1\% \gray{(0.11x)} & 11.1\% \gray{(0.11x)} \\
ASPKD         & 0 & \checkmark & 39.2\% \gray{(0.40x)} & 39.0\% \gray{(0.40x)} & 35.4\% \gray{(0.36x)} & 32.0\% \gray{(0.33x)} \\

Real Guidance & 1 & \checkmark & 51.2\% \gray{(0.52x)} & 52.0\% \gray{(0.53x)} & 43.9\% \gray{(0.45x)} & 40.6\% \gray{(0.41x)} \\

DA-Fusion     & 1 & \ding{55} & 59.0\% \gray{(0.60x)} & 53.3\% \gray{(0.54x)} & 58.8\% \gray{(0.60x)} & 48.6\% \gray{(0.50x)} \\

Stealix (ours)    & 1 & \ding{55} & \textbf{65.9\%} \gray{(0.67x)} & \textbf{67.9\%} \gray{(0.69x)} & \textbf{66.0\%} \gray{(0.67x)} & \textbf{51.9\%} \gray{(0.53x)} \\
\bottomrule
\end{tabular}%
}
\end{center}
\end{table}

\section{Different Victim Model Architectures}
\label{app:victim_architecture}

We analyze the performance of Stealix across different victim model architectures on EuroSAT, including ResNet18, ResNet34, VGG16, and MobileNet, as shown in~\Cref{tab:victim_architecture}.
Using ResNet18 as the attacker architecture, Stealix consistently performs well across these architectures, demonstrating its robustness to variations in the victim model. The ability to generalize across diverse architectures highlights the adaptability and effectiveness of Stealix in real-world scenarios where the attacker may not know the exact architecture of the victim model.

\begin{table}[h!]
\caption{Performance comparison of Stealix against different victim architectures (ResNet18, ResNet34, VGG16, MobileNet) with the attacker model architecture set to ResNet18 across all experiments on EuroSAT.}
\vskip 0.1in
\begin{small}
\begin{center}
\label{tab:victim_architecture}
\begin{tabular}{c|cccc}
\toprule
\multirow{2}{*}{\textbf{Method}} & \multicolumn{4}{c}{\textbf{Victim architecture}} \\ 
\cmidrule(lr){2-5}
 & \textbf{ResNet18} & \textbf{ResNet34} & \textbf{VGG16} & \textbf{MobileNet} \\ 
\midrule
Victim & 
98.4\% \textcolor{gray}{(1.00x)} & 
98.2\% \textcolor{gray}{(1.00x)} & 
98.2\% \textcolor{gray}{(1.00x)} & 
96.9\% \textcolor{gray}{(1.00x)} \\ 

Stealix (ResNet18) & 
66.2\% \textcolor{gray}{(0.67x)} & 
65.9\% \textcolor{gray}{(0.67x)} & 
73.4\% \textcolor{gray}{(0.75x)} & 
66.0\% \textcolor{gray}{(0.68x)} \\
\bottomrule
\end{tabular}%
\end{center}
\end{small}
\end{table}

\section{Limitations of DFME}
\label{app:limitatios_of_dfme}
We analyze the performance of DFME~\citep{dfme} under realistic attack scenarios.
Following the original DFME setup, we attempted to extract our ResNet34 victim model trained on CIFAR-10 using 2 million queries per class with soft-label access, achieving an attacker model accuracy of 87.4\%, which is comparable to the results reported in the original work. However, DFME generates images with pixel values in the range $(-1, 1)$ due to the use of Tanh activation, which is incompatible with real-world APIs that expect standard image formats (e.g., pixel values in $[0, 255]$). After quantizing these images to the standard format, the attacker model accuracy dropped to 76.4\%, despite using the same query budget. This performance degradation occurs because DFME relies on adding small perturbations to the generated images to estimate gradients via forward differences~\citep{forward_differences}. Quantization can negate these subtle perturbations. Furthermore, when the victim model provides only hard-label outputs as a defense mechanism, the attacker model accuracy further decreased to 23.7\%. In this case, the output labels remain constant under small input perturbations, rendering forward difference methods ineffective for gradient estimation and significantly limiting the attacker's ability to train the generator.

We present the results across all datasets in~\Cref{tab:DFME}. 
In the case of PASCAL, we reduced the batch size from 256 to 64 due to computational constraints imposed by the large image size ($256 \times 256$ pixels). Notably, DFME fails to extract the PASCAL victim model, likely due to this higher image resolution. Furthermore, for the fine-grained EuroSAT dataset, even with soft-label access and without quantization, the attacker model achieves only 19.0\% accuracy.

\begin{table}[h!]
\caption{Performance of DFME on various datasets under different settings with a query budget of 2M per class. Victim model accuracies are provided for reference.}
\label{tab:DFME}
\vskip 0.1in
\begin{small}
\begin{center}
\begin{tabular}{lcccc}
\toprule
\multicolumn{1}{l}{\bf Method}  & \multicolumn{1}{c}{\bf EuroSAT} & \multicolumn{1}{c}{\bf PASCAL} & \multicolumn{1}{c}{\bf CIFAR10} & \multicolumn{1}{c}{\bf DomainNet} \\ \midrule

Victim        & 98.2\% \gray{(1.00x)} & 82.7\% \gray{(1.00x)}  & 93.8\% \gray{(1.00x)}  & 83.9\% \gray{(1.00x)}  \\
 \midrule

DFME      & 19.0\% \gray{(0.19x)} & 6.6\% \gray{(0.08x)} & 87.4\% \gray{(0.93x)} & 83.0\% \gray{(0.99x)} \\
 \text{ + Quantization}     & 10.2\% \gray{(0.10x)} & 6.6\% \gray{(0.08x)} & 76.4\% \gray{(0.81x)} & 72.0\% \gray{(0.86x)} \\
 \text{ + Hard label}     & 11.1\% \gray{(0.11x)} & 6.6\% \gray{(0.08x)} & 23.7\% \gray{(0.25x)} & 18.5\% \gray{(0.22x)} \\
 
\bottomrule
\end{tabular}%
\end{center}
\end{small}
\end{table}


\section{Stealix with Soft Labels}
\label{app:soft_label}

\begin{figure}[ht!]
    \centering
    \includegraphics[width=0.85\textwidth]{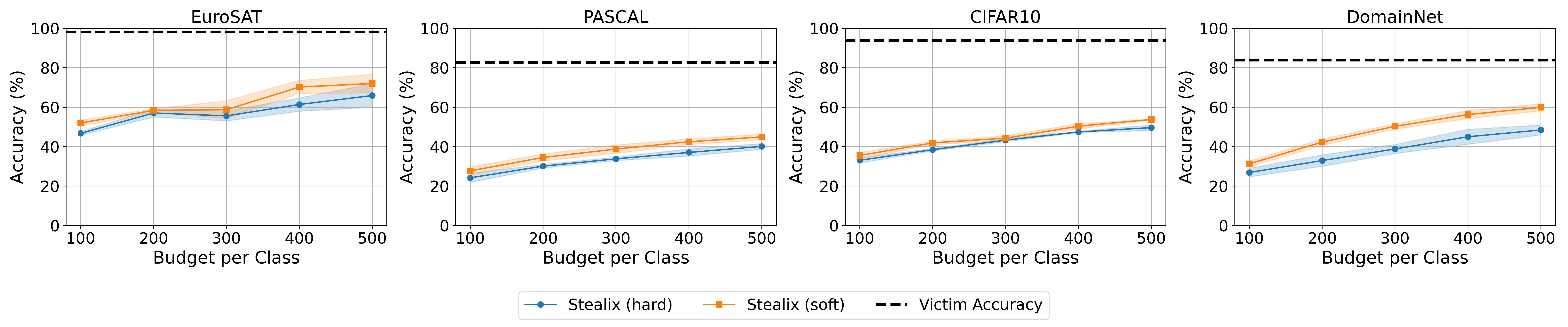}
    \vskip 0.1in
    \caption{Performance comparison of Stealix with hard label and soft label access across EuroSAT, PASCAL, CIFAR-10, and DomainNet at varying query budgets.}
    \label{fig:hard_soft_label}
\end{figure}

In this experiment, we evaluate the impact of soft-label access on the attacker model accuracy compared to the hard-label-only scenario. Since Stealix's prompt evolution only relies on hard labels for calculating prompt consistency, the same synthetic images are used to train the attacker model under both conditions, with the only difference being whether the labels are hard or soft (full probability predictions). As shown in~\Cref{fig:hard_soft_label}, Stealix consistently achieves higher accuracy with soft-label access across all datasets, as soft labels provide richer information through confidence scores, resulting in improved model performance. This underscores the importance of defenses like hard-label-only outputs to limit the effectiveness of model stealing attacks.
However, hard-label defenses merely slow down the attack, increasing the required query budget without fully preventing model theft. Given the high quality and alignment of synthetic images with the victim's data, the attack remains viable over time. This highlights the need for more advanced defense strategies to better address this threat in future research.

\section{DA-Fusion as Data Augmentation}
\label{app:model_stealing_outperforms_data_augmentation}
Having one image per class is a realistic setup and differs from having full access to victim data or its distribution. This reflects real-world threats posed by competitors in the same field, aiming to provide similar services. 
While attackers can use DA-Fusion to augment the seed images to train the attacker model without querying the victim model, we demonstrate that model stealing still provides a substantial performance improvement.
We compare the accuracy of attacker models under a model stealing setup versus a data augmentation setup, with a query budget of 500 per class. \Cref{tab:da_fusion_as_data_augmentation} shows that performance degrades significantly with DA-Fusion when relying solely on class labels for training instead of using predictions from the victim model, highlighting that model stealing is essential, even with one image per class.

\begin{table}[h!]
\caption{Comparison of attacker model training with and without victim queries, showing accuracy with a 500-query budget per class; DFME uses 2M.}
\label{tab:da_fusion_as_data_augmentation}
\vskip 0.1in
\addtolength{\tabcolsep}{-0.2em}
\begin{center}
\begin{small}
{\fontsize{8}{9}\selectfont  
\begin{tabular}{ccccccc}
\toprule
\multicolumn{1}{c}{\bf Method} & \multicolumn{1}{c}{\bf Query victim} & \multicolumn{1}{c}{\bf EuroSAT} & \multicolumn{1}{c}{\bf PASCAL} & \multicolumn{1}{c}{\bf CIFAR10} & \multicolumn{1}{c}{\bf DomainNet} \\ \midrule

Victim        & - & 98.2\% \gray{(1.00x)} & 82.7\% \gray{(1.00x)}  & 93.8\% \gray{(1.00x)}  & 83.9\% \gray{(1.00x)}  \\ \midrule

Stealix (ours)    & \ding{51} & \textbf{65.9\%} \gray{(0.67x)} & \textbf{40.0\%} \gray{(0.48x)} & \textbf{49.6\%} \gray{(0.53x)} & \textbf{48.4\%} \gray{(0.58x)} \\
DA-Fusion    & \ding{51} & 59.0\% \gray{(0.60x)} & 16.4\% \gray{(0.20x)} & 26.7\% \gray{(0.28x)} & 28.4\% \gray{(0.34x)} \\
DA-Fusion     & \ding{55} & 29.9\% \gray{(0.30x)} & 10.7\% \gray{(0.13x)} & 18.9\% \gray{(0.20x)} & 17.9\% \gray{(0.21x)} \\

\bottomrule
\end{tabular}
}
\end{small}
\end{center}
\end{table}

\newpage
\section{Limited Medical Knowledge}
\label{app:medical_image_model_stealing}
As generative priors like diffusion models are trained on publicly available data, the absence or limited presence of domain-specific knowledge, such as medical expertise, would have impact on the performance of model stealing relies on these models. However, this issue applies universally to all model stealing methods that rely on diffusion models, not specifically to ours. Our experiment results in \Cref{tab:performance-table} show that diffusion models can be leveraged more effectively in model stealing when they describe the data well but are not properly prompted. In other words, \textbf{our approach shares the same lower-bound as existing methods but significantly improves the upper-bound}, achieving an approximate 7–22\% improvement compared to the second-best method, as shown in \Cref{tab:performance-table}.

\begin{figure}[ht!]
    \vskip 0.1in
    \begin{center}
    \includegraphics[width=0.75\textwidth]{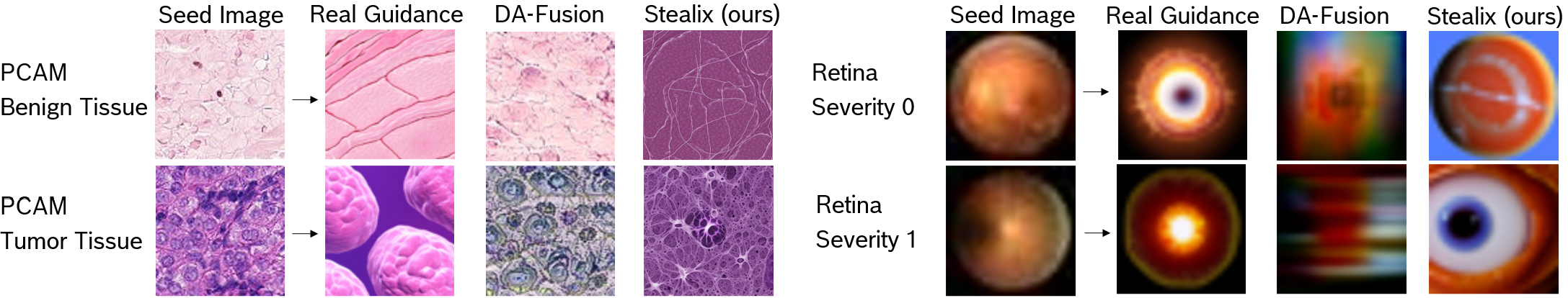}
    \vskip 0.1in
    \caption{Qualitative comparison of images generated by Real Guidance, DA-Fusion, and Stealix on the PCAM and RetinaMNIST datasets. Other baselines include: Knockoff uses CIFAR10 as query data, DFME synthesizes noise images, and ASPKD uses the same images as Real Guidance.}
    \label{fig:qualitative_medical}
    \end{center}
    \vskip 0.1in
\end{figure}

With that being said, we conducted an experiment analyzing performance when diffusion models have limited domain-specific knowledge. We consider two medical datasets: PatchCamelyon (PCAM)~\citep{veeling2018rotation} and RetinaMNIST~\citep{yang2023medmnist}. In PCAM, class names are ``benign tissue'' and ``tumor tissue''. RetinaMNIST involves a five-level grading system for diabetic retinopathy severity, with class names as ``diabetic retinopathy $i$,'' where $i$ ranges from 0 to 4 for severity. We conduct experiments using three random seeds and report the mean attacker accuracy below, following the setup described in \Cref{subsec:experimental_setup}. The victim model uses the ResNet34 architecture, while the attacker model is based on ResNet18. The qualitative comparison in \Cref{fig:qualitative_medical} shows that the diffusion model struggles to synthesize Retina-like images, highlighting its limited knowledge. However, the results in \Cref{tab:performance-table-medical} show that methods with generative priors still outperform Knockoff Nets and DFME, confirming the value of priors, though the improvements decrease as the data deviates from the diffusion model's distribution, resulting in only modest gains of Stealix in such cases.

In summary, our approach provides (1) significant improvement when diffusion models can describe the data and (2) comparable or slightly better performance when they have limited domain knowledge.

\begin{table}[ht!]
\caption{Attacker model accuracy for medical dataset with a query budget of 500 per class; DFME uses 2M.}
\label{tab:performance-table-medical}
\vskip 0.1in
\addtolength{\tabcolsep}{-0.2em}
\begin{center}
\begin{small}
\begin{tabular}{ccccccc}
\toprule
\multicolumn{1}{c}{\bf Method} & \multicolumn{1}{c}{\bf \#Seed images}& \multicolumn{1}{c}{\bf Class name}  & \multicolumn{1}{c}{\bf PCAM} & \multicolumn{1}{c}{\bf RetinaMNIST} \\ \midrule

Victim        & - & - & 91.2\% \gray{(1.00x)} & 61.7\% \gray{(1.00x)}  \\

(KD)            & - & - & 76.3\% \gray{(0.84x)} & 59.4\% \gray{(0.96x)} \\ \midrule

Knockoff      & 0 & \ding{55} & 50.0\% \gray{(0.55x)} & 56.1\% \gray{(0.91x)} \\
DFME          & 0 & \ding{55} & 50.0\% \gray{(0.55x)} & 46.1\% \gray{(0.75x)} \\
ASPKD         & 0 & \checkmark & 60.1\% \gray{(0.66x)} & 55.3\% \gray{(0.90x)} \\

Real Guidance & 1 & \checkmark & 61.8\% \gray{(0.68x)} & 56.1\% \gray{(0.91x)} \\

DA-Fusion     & 1 & \ding{55} & 61.5\% \gray{(0.68x)} & 56.7\% \gray{(0.92x)} \\

Stealix (ours) & 1 & \ding{55} & \textbf{62.2\%} \gray{(0.68x)} & \textbf{58.0\%} \gray{(0.94x)} \\
\bottomrule
\end{tabular}
\end{small}
\end{center}
\end{table}



\end{document}